\documentclass[preprint2]{aastex}
\usepackage{psfig}
\begin{document}
  \title{\large Observational manifestations of solar magneto-convection ---
   center-to-limb variation}
\author{Mats Carlsson\altaffilmark{1}}
\affil{Institute of Theoretical Astrophysics, University of Oslo,
P.O.~Box~1029, Blindern, N--0315 Oslo, Norway}
\email{mats.carlsson@astro.uio.no}
\altaffiltext{1}{also at Center of Mathematics for Applications, University of Oslo,
P.O.~Box~1053, Blindern, N--0316 Oslo, Norway}
\author{Robert F.~Stein}
\affil{Department of Physics \& Astronomy,
 Michigan State University,
 East Lansing, MI 48824, USA}
\email{bob@steinr.pa.msu.edu}
\author{{\AA}ke Nordlund}
\affil{NBIfAFG, University of Copenhagen, DK-2100 Copenhagen, Denmark}
\email{aake@astro.ku.dk}
\and
\author{G\"oran B. Scharmer}
\affil{The Institute for Solar Physics of the Royal Swedish Academy of
  Sciences, Alba Nova University Center, SE-10691, Stockholm, Sweden}
\email{scharmer@astro.su.se}

\begin{abstract}
We present the first center-to-limb G-band images
synthesized from high resolution simulations of solar magneto-convection. 
Towards the limb the simulations show
``hilly" granulation with dark bands on the far side, 
bright granulation walls and striated faculae, similar to observations. 
At disk center G-band bright points are flanked by dark lanes.
The increased brightness in magnetic elements is due to their
lower density compared with the surrounding intergranular medium. One
thus sees deeper layers where the temperature is higher. At a given
geometric height, the magnetic elements are cooler than the
surrounding medium. In the G-band, the contrast is further increased by
the destruction of CH in the low density magnetic elements. The optical
depth unity surface is very corrugated.  Bright granules have their
continuum optical depth unity 80 km above the mean surface, the magnetic
elements 200-300 km below.  
The horizontal temperature gradient is especially large next to flux
concentrations.  When viewed at an angle, the deep magnetic
elements optical surface is hidden by the granules and the bright points
are no longer visible, except where the ``magnetic valleys" are aligned 
with the line of sight.  Towards the limb, the low
density in the strong magnetic elements causes unit line-of-sight
optical depth to occur deeper in the granule walls behind than for rays
not going through magnetic elements and variations in the field strength
produce a striated appearance in the bright granule walls. 
\bigskip
\end{abstract} 


\section{Introduction}
%
G-band filtergram bright points are commonly used as proxies for small
scale magnetic flux concentrations, sometimes called ``flux tubes'',
\citep{Muller+Roudier1984,Berger+Schrijver+Shine+etal1995,Berger+Title2001}, 
although it is known that not
all magnetic elements are associated with bright points
\citep{Berger+Title1996}.  Such bright points are also observed in the
photospheric continuum, but with lower contrast \citep{Dunn+Zirker1973}.
The reason for this association is the decreased opacity in
the magnetic concentration
\citep{Kiselman+Rutten+Plez2001,Rutten+Kiselman+RouppevanderVoort+Plez2001,
SanchezAlmeida+AsensioRamos+etal2001,%
Steiner+Hauschildt+Bruls2001,%
Uitenbroek2003,%
Keller+Schussler+Vogler+Zakharov2004}.  The
primary reason for the lower opacity is the lower density in magnetic
concentrations, which maintain rough pressure equilibrium with their
surroundings.  The contrast in the G-band is increased by the
destruction of the CH molecules, again due to the low density,
although some have also cited increased temperatures as a cause
\citep{Schussler+Shelyag+Berdyugina+etal2003,%
SanchezAlmeida+AsensioRamos+etal2001,%
Steiner+Hauschildt+Bruls2001}.
Towards the limb magnetic concentrations are visible as bright faculae,
where one sees the hot granule walls behind the low
opacity magnetic concentrations 
\citep{Spruit1976,Spruit1977,Lites+Scharmer+Berger+Title2004,
Keller+Schussler+Vogler+Zakharov2004}.

Here we examine the variation in the appearance of the G-band from
disk center toward the limb using images calculated from a solar
magneto-convection simulation.  

\section{Methods} \label{methods}

%

The 3D magneto-convection simulations include LTE ionization and
excitation in the equation of state, and non-grey, LTE radiation
transfer in the energy balance.
They cover a small region of 6~$\times$~6~Mm,
sufficient to include one meso-granule and many granules, and extend
from the temperature minimum down to a depth of 2.5~Mm below 
$<\tau_{\mbox{500}}>=1$.  The resolution is 25~km horizontally and
varies from 15~km in the upper layers to 35~km vertically in the lower layers  
\citep{Stein+Nordlund1998}.
The simulation
is started with a uniform vertical magnetic field of 250 G superimposed
on a snapshot of non-magnetic convection and then allowed to relax.  The
magnetic field is quickly swept to the boundaries of the deep lying
mesogranules and concentrated to typical strengths 1.7 kG at the level
$<\tau_{\mbox{500}}>=1$.

%

The emergent spectrum was calculated in LTE in two
wavelength intervals of 3~nm width centered on 430.68~nm (G-band) and
436.52~nm (G-continuum) (all wavelengths given as vacuum
wavelengths). Molecular equilibrium was calculated including all
di-atomic molecules of hydrogen, carbon, nitrogen and oxygen.  In each
wavelength band, 1364 frequency points were used for the spectrum
calculation. 
Line opacities from 21 atomic species and molecular lines from CH
were included with data from \citet{Kurucz+Bell1995} and 
\citet{Kurucz1993}. Depth-dependent Voigt profiles were used for all
lines. In total 1838 lines were considered.
The calculated emergent intensities were
multiplied with filter transmission curves 
from the Swedish 1-meter Solar Telescope filters with central
wavelengths as above and FWHM of 1.08~nm (G-band) and 1.15~nm
(G-continuum) and integrated over
wavelength to create synthetic images in the G-band and
G-continuum. 


\section{Simulated Observations}

We investigate the response of the G-band and G-continuum to magnetic
concentrations, the so-called ``flux tubes'', at disk center and
towards the limb.  We consider disk center first.

\subsection{Disk Center} \label{center}
%
Hydrodynamic
calculations using the same code have previously been checked by comparing simulated
and observed iron line profiles \citep{Asplund+Nordlund+Trampedach+etal2000}.  
The accuracy of both the current simulation and the G-band calculation was
checked by comparing the spatially averaged G-band spectrum calculated
from a simulation snapshot with observations.  The simulated
G-band spectrum calculated without any micro- or macro-turbulence
or extra wing damping is in excellent agreement with the Jungfraujoch
atlas \citep{solaratlas},
with only a 8\% rms difference between the two.  These tests verify
the realism of the simulation and the accuracy of the G-band
calculation.

\begin{figure}
\centerline{\psfig{file=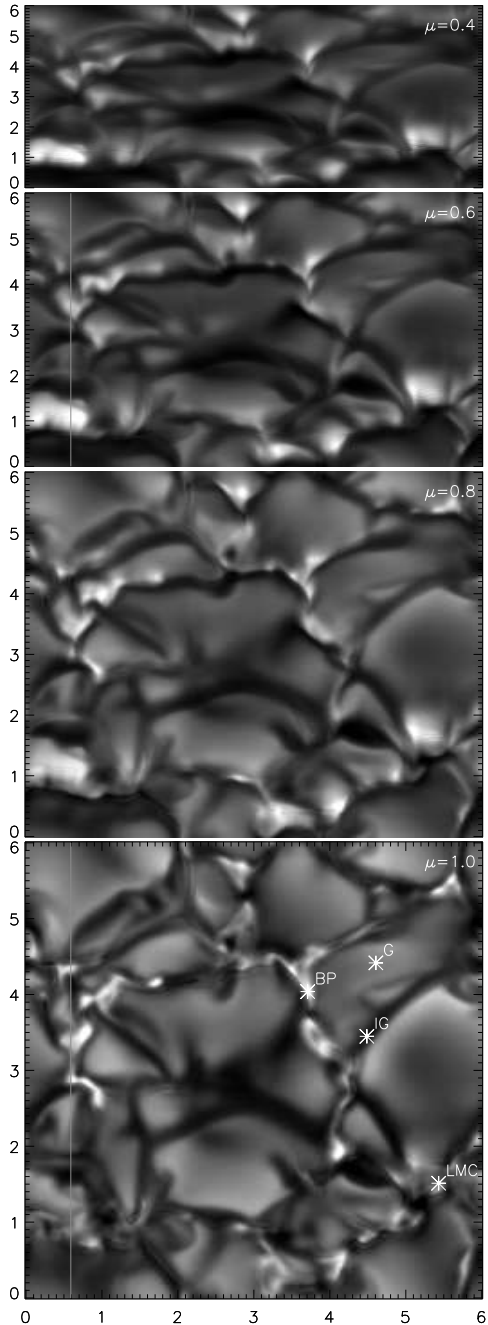,width=7cm}}
\caption{G-band images at disk center and $\mu=0.8, 0.6, 0.4$. The
bright line shows the slice along which we show the atmospheric
structure and location of peak emission.  Axes are distances in Mm.
The panels have individual scalings.  Points marked G, IG, BP and
LMC refer to the locations where the temperature stratification is shown
in figure \ref{figtz}.
\label{ctr-limb-im}}
\end{figure}
\begin{figure}[t]
\centerline{\psfig{file=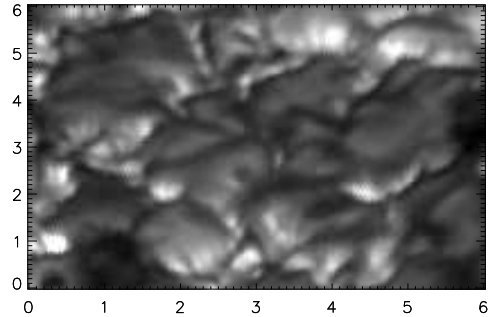,width=7cm}}
\caption{Observed G-band intensity from the Swedish 1-m Solar Telescope
at $\mu=0.63$. Data processing done by Luc Rouppe van der Voort.
\label{ctr-limb-obs}}
\end{figure}

Figure \ref{ctr-limb-im} shows the G-band images
at $\mu$ = 1.0, 0.8, 0.6 and 0.4 for
a snapshot from our simulation.  The appearance is very similar
in its general properties to observed images (Fig.~\ref{ctr-limb-obs}).
Our average field strength of 250 G is more like a plage region,
but the appearance is more like the quiet Sun.  With higher 
spatial resolution in the simulations the same appearance would occur
for smaller average fields, because the simulations would then allow
compression to smaller scales. The same typical field strengths of 1.7
kG would then occur but in structures of smaller spatial extent.

\begin{figure}[!ht]
\centerline{\psfig{file=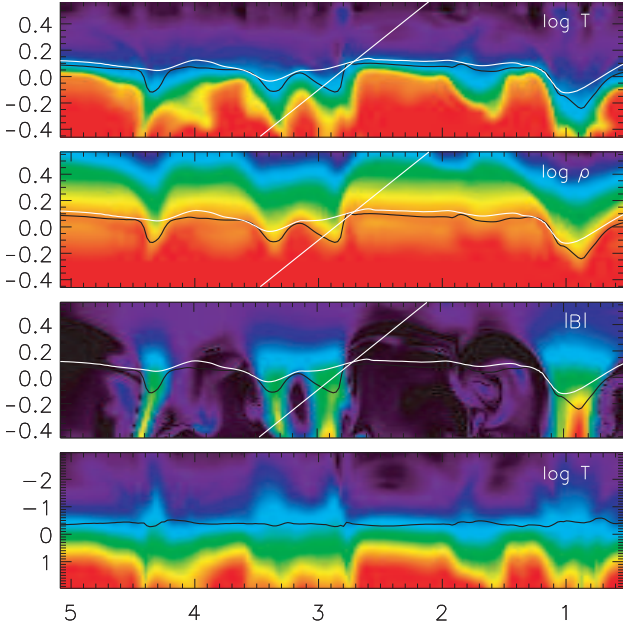}}
\caption{Temperature, density and magnetic field strength along a
vertical slice through magnetic and non-magnetic regions, with the
average formation height for the G-band intensity for a vertical
ray (black line) and at $\mu=0.6$ (white line). One ray at $\mu=0.6$
is also shown in white. Axes are distances in
Mm. The bottom panel shows temperature as function of lg $\tau_{500}$.
\label{slice}}
\end{figure}

\begin{figure}[!ht]
\centerline{\psfig{file=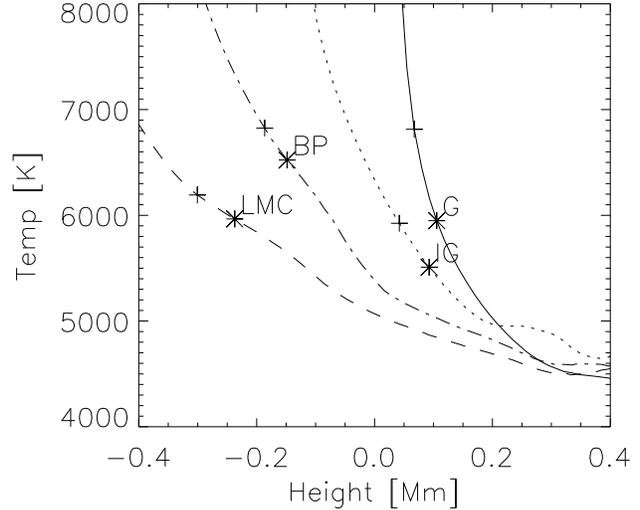}}
\caption{Temperature vs. height for four locations: G=granule,
IG=intergranular lane, BP=magnetic bright point, LMC=large magnetic
flux concentration.  Stars are mean formation height of the G-band 
and pluses of the G-continuum.  Magnetic flux concentrations are cooler
than their surroundings at a given geometric height but their radiation 
emerges from deeper, hotter layers.
\label{figtz}}
\end{figure}

The magnetic concentrations are partially evacuated and cooler
than their surroundings at a given geometric layer, and
in approximate pressure equilibrium
with the surrounding non-magnetic plasma (Fig.~\ref{slice}). 
Because they are cooler and less dense, the layer from which radiation 
emerges lies deeper, often sufficiently deep that the temperature
is higher than at the formation height in non-magnetic granules  
(Fig. \ref{figtz}).  
The $\tau_{500}$=1 surface is extremely corrugated, ranging from
97 km above (in granules) to 323 km below (Wilson depression) the average.
Higher temperatures at
$\tau_{500}$=1 are due to radiative heating from the hot walls of the magnetic
concentrations as predicted by \citet{Spruit1977,Spruit1976}.
Indeed, in calculations where horizontal radiative transfer is
omitted, these higher temperatures at $\tau_{500}$=1 are absent,
confirming the role of radiative heating from hot side walls.  As
a result of this heating, the magnetic concentrations often appear
as bright points in the intergranular lanes.  This is especially
apparent in the G-band, where the contrast is enhanced because
CH is destroyed in the low density magnetic concentrations.  
All such bright points are associated with strong magnetic
fields.  However, there are significant numbers of locations of similar
strong field that are not associated with G-band brightness at disk
center (e.g. fig. \ref{ctr-limb-im} lower right and left).  
Whether a concentration appears bright or dark depends on
its size -- larger concentrations tend to be dark while smaller
ones tend to be bright \citep{Spruit1977}.

The bottom panel of Fig~\ref{slice} shows
the temperature as function of lg $\tau_{500}$. The contrast
in temperature between magnetic concentrations and non-magnetic areas 
increases with decreasing optical depth giving larger intensity contrast
with increasing opacity (e.g. Ca H,K). The G-band has its mean formation 
height (black
line in bottom panel) at lg $\tau_{500}$=-0.48 corresponding to a mean
formation height 54~km above where $\tau_{500}$=1, therefore giving a larger
contrast than in the continuum. The contrast enhancement by the destruction
of CH is seen as a dip in the curve showing the mean formation optical
depth in the bottom panel. Note also that the G-band intensity has its
peak contribution at similar heights as the continuum (that's why the 
granulation pattern looks similar) but the many spectral lines give a broad
tail of the contribution function to higher layers. This causes a more
``fuzzy'' image than the sharper contribution function from a monochromatic
diagnostic would give.

One often sees especially dark lanes along the sides of the G-band
bright points.  These are the cool intergranular downflows that have
been excluded from the magnetic concentrations.  In addition,
near the layer of $<\tau_{500}>=1$, the magnetic concentrations
are surrounded by rings of slightly lower temperature 
where the magnetic field strength gradient is largest.

Finally, while most
of the granulation appears normal, where there is a wide, strong
magnetic field concentration along the bottom edge of the image, the
granulation appears very disturbed, with lots of small, irregular
bright and dark features.

\subsection{Center to Limb Variation} \label{limb}
%
As one observes towards the limb several differences in the surface
appearance occur (Fig.~\ref{ctr-limb-im}).  First, the granules
develop a three-dimensional pillow appearance with the granules
higher than the intergranular lanes, and their near sides brighter
than their far sides.  This is partly due to the true three-dimensional
structure of granulation with the granules observed higher than the
intergranular lanes.  Partly it is due to the near side being seen
more normal to the line of sight and the far side at a more glancing
angle.  Finally, this is enhanced by the magnetic concentrations,
which because of their low density and resulting low opacity allow
one to see deeper into the hot granules behind them (the ``hot
wall'' effect 
\citep{Spruit1977,Spruit1976,Keller+Schussler+Vogler+Zakharov2004}).  
This is further enhanced by the occurrence of the
largest horizontal temperature gradients along the boundaries of
some of the magnetic concentrations.  These features are illustrated
in the figure \ref{slice} which show the temperature, density,
magnetic field strength and the
mean formation height of the G-band intensity for both a
vertical line of sight and one at $\mu=0.6$, in a vertical slice through
the simulation domain (location shown in fig. \ref{ctr-limb-im} as light
vertical line).
Note that the mean formation height for a slanted ray is more constant
than for a vertical ray. The deep, low density magnetic concentrations
get hidden by the fact that the ray goes through higher density material
above the neighboring granule.

In figure \ref{ctr-limb-im} the bright faculae on the left and right 
sides near the bottom (at $\mu=0.8, 0.6$) are
behind the largest flux concentration at this time, which appear
mostly dark in the disk center image.  As one looks further toward
the limb some of the faculae (especially in the upper part
of the image) disappear because they lie deep in the intergranular
lanes and become hidden behind the higher intervening granule tops.
Where, however, the bright walls are viewed along an intergranular
lane or where the magnetic concentration is sufficiently wide they 
remain clearly visible.

Another common phenomenon in observations and the simulated images
is the appearance of striations in the bright granule walls.  These
striations are caused by variations in the magnetic field strength
in front of the hot granule walls.  Where the field is weaker, the
density is higher, so the opacity larger.  This effect is enhanced
by a higher CH concentration also due to the higher density.  Thus,
where the magnetic field is weaker, the radiation emerges from
higher, cooler layers, so these locations appear darker.


\acknowledgements
This work was supported by the Danish Center for Scientific
Computing, by the Research Council of Norway
grant 146467/420 and a grant of computing time
from the Program for Supercomputing, by NASA grants
NAG 5 12450 and NNG04GB92G and by NSF
grant AST0205500. The Swedish 1m Solar Telescope is operated on the
island of La Palma by the Institute for Solar Physics of the Royal
Swedish Academy of Sciences in the Spanish Observatorio del
Roque de los Muchachos of the Instituto de Astrof{\i}sica de Canarias.
Elin Marthinussen is thanked for providing line synthesis
atomic and molecular data.

\end{document}